\documentclass[lettersize,journal]{IEEEtran}
\usepackage{amsmath,amsfonts}
\usepackage{algorithmic}
\usepackage{array}
\usepackage[caption=false,font=normalsize,labelfont=sf,textfont=sf]{subfig}
\usepackage{textcomp}
\usepackage{stfloats}
\usepackage{url}
\usepackage{graphicx}
\usepackage{algorithm}
\def\BibTeX{{\rm B\kern-.05em{\sc i\kern-.025em b}\kern-.08em
    T\kern-.1667em\lower.7ex\hbox{E}\kern-.125emX}}
\usepackage{balance}
\DeclareMathOperator*{\argmax}{argmax}
\begin{document}

\title{Power Control of Multi-Layer Repeater Networks (POLARNet)}

\author{Johan Siwerson, Johan Thunberg
\thanks{This work has been submitted to the IEEE for possible publication. Copyright may be transferred without notice, after which this version may no longer be accessible.}
\thanks{The authors are with the Department of Electrical and Information Technology (EIT), Lund University, SE-22100 Lund, Sweden (e-mail: johan.siwerson@eit.lth.se; johan.thunberg@eit.lth.se).}
\thanks{This work was partially supported by the Wallenberg AI, Autonomous Systems and Software Program (WASP) funded by the Knut and Alice Wallenberg Foundation.}
}

\maketitle
\begin{abstract}
In this letter we introduce POLARNet -- power control of multi-layer repeater networks -- for local optimization of SNR given different repeater power constraints. We assume relays or repeaters in groups or layers spatially separated. Under ideal circumstances SISO narrow-band communication and TDD, the system may be viewed as a dual to a deep neural network, where activations, corresponding to repeater amplifications, are optimized and weight matrices, corresponding to channel matrices, are static. Repeater amplifications are locally optimized layer-by-layer in a forward-backward manner over compact sets. The method is applicable for a wide range of constraints on within-layer power/energy utilization, is furthermore gradient-free, step-size-free, and has proven monotonicity in the objective. Numerical simulations show significant improvement compared to upper bounds on the expected SNR. In addition, power distribution over multiple repeaters is shown to be superior to optimal selection of single repeaters in the layers.  
\end{abstract}


\section{Introduction}
\label{sec:introduction}
\IEEEPARstart{T}{o} meet increasing demands on coverage and capacity, network-controlled repeaters have been suggested for network coverage extension~\cite{3gpp_ncr, wen2024shaping} to complement or augment D-MIMO~\cite{spectral_eff_D_mimo,haliloglu2023distributed} and work alongside technologies such as large intelligent surfaces~\cite{hu2018beyond}. By integrating repeaters into a D-MIMO framework, repeater assisted MIMO (RA-MIMO) may enhance channel diversity and coverage while maintaining the benefits of spatial multiplexing~\cite{saras_paper}. Repeaters, which amplify and forward signals, could be strategically deployed to extend the reach at a low cost and complexity in a distributed network.

In such amplify-and-forward relay systems, optimization of amplifications~\cite{sanguinetti2012tutorial} is key to accomplishing various objectives. Recently, a max-min amplification method was proposed for RA-MIMO and compared to cell-free Massive MIMO (cfmMIMO) in terms of user fairness and repeater energy efficiency~\cite{topal2025fair}. In~\cite{tsai2010capacity} capacity scaling was investigated in repeater-aided MIMO systems in line-of-sight (LOS) environments. In addition to the optimization objectives for repeater amplification/control mentioned above, these settings present further challenges, one of which is ensuring network stability in the presence of feedback~\cite{larsson_stability}.

\begin{figure}
    \centering
    \includegraphics[width=0.4\textwidth]{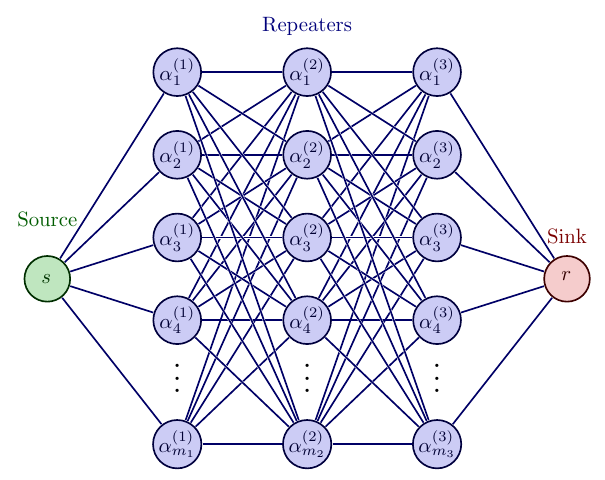}
    \caption{System model of a multi-layer repeater network with a single base station (green), three repeater layers (purple) and a single user (red).}
    \label{fig:dag_model}
\end{figure}

In this letter we propose POLARNet — power control of multi-layer repeater networks. We consider simple relays or repeaters in groups spatially separated. If the repeaters in each group comprise a line formation, they can be seen as forming a layer, where the repeater amplifications (or gains) can be controlled or selected via local optimization to improve a common objective such as SNR. 

Such line formations of repeaters have earlier been considered \emph{e.g.} in the context of increasing cellular network energy efficiency for railway corridors~\cite{schumacher2022increasing}, where information is relayed along the line of repeaters. Here instead, repeaters in a layer receive information from a previous layer and then amplify the signal for reception at a subsequent layer. This setting is illustrated in Fig.~\ref{fig:dag_model} where three layers with varying amount of repeater nodes are shown in the middle (purple),  base station BS node is shown on the left (green) and UE-nodes are shown on the right (red). 

There is a plethora of methods for solving optimization problems within the context of communication systems~\cite{liu2024survey}. However, in this letter, we introduce a new local optimization method for SNR, which is tailored to the setting at hand. 
Under ideal assumptions of narrowband communication and time division duplex (TDD) and communication between consecutive layers only, the system may be viewed as a complex-valued dual to a deep neural network. 
Positive repeater-gains correspond to activation functions and shall be locally optimized within compact sets we refer to as activation sets. Repeater amplifications are locally optimized iteratively layer-by-layer in a forward-backward manner similar to backpropagation for deep neural networks~\cite{rumelhart1986learning}. 

The introduced method POLARNet improves the objective function in each iteration is gradient-free, step-size-free, and allows for optimization over non-smooth and even discrete constraint sets. Numerical simulations show significant improvement compared to upper bounds on the expected SNR for random power allocation. In addition, power distribution over multiple repeaters is shown to be superior to optimal selection of single repeaters in the layers.


\section{System Model}
\label{subsec:system_model}
Repeater nodes (in the multi-layer repeater networks introduced in Section~\ref{sec:introduction}) are viewed as full-duplex reciprocal relays, instantaneously (or with a negligible delay) amplifying and forwarding signals. We refer to \emph{e.g.}~\cite{saras_paper} for more technical details on implementation strategies in more general scenarios. We further assume communication is directional and only occurs between consecutive repeater layers using TDD for frequency non-selective and slowly fading channels. 

Our system model contains a base station (BS) serving a single user equipment (UE) via interconnected layers of repeaters. The structure of the multi-layer repeater network may be captured by a  directed acyclic graph (DAG) for downlink (DL) and the reversed such DAG-graph for uplink (UL). There are $n$ layers of repeater nodes separating transmitter(s) and receiver(s), see Fig.~\ref{fig:dag_model}, where DL is visualized. Each layer of repeaters $i$ comprises a group of $m_i$ static repeaters indexed by the number of hops from the BS in DL. In Fig.~\ref{fig:dag_model} repeater locations are collinear, which is intuitively favorable to mitigate within-layer interference. We consider the network in isolation without external dependencies and interference.  

In the network, the $j$'th repeater node in layer $i$ has a corresponding weight $\alpha^{(i)}_j \in \mathbb{R}^+$, comprising its amplification or gain. Throughout, $\mathbb{R}^+$ denotes the non-negative real numbers. The vector $\bold{H}_{1,0} \in \mathbb{C}^{m_{1} \times 1}$ contains the channels between BS and repeater layer 1; the matrix $\bold{H}_{i+1,i} \in \mathbb{C}^{m_{i+1} \times m_{i}}$, for $i \in \{1,2, \ldots, n-1\}$, contains the pairwise channels between repeater layer $i$ and repeater layer $i+1$; and the vector $\bold{H}_{n+1,n} \in \mathbb{C}^{1 \times m_n}$ contains the channels between repeater layer $n$ and the UE. We assume these channels to be known. An in-depth study on for example sufficient conditions for channel estimation is out scope for this letter. 

The total SISO channel ${h}_{\text{tot}} \in \mathbb{C}$, or simply the channel for transmission between BS and UE, is given by a  chain of matrix products 
\begin{equation}
\label{eq:h_tot}
    {h}_{\text{tot}} = \bold{H}_{n+1,n} \bold{D}_{n}\bold{H}_{n,n-1} \bold{D}_{n-1} \cdots \bold{H}_{3,2} \bold{D}_2 \bold{H}_{2,1}\bold{D}_1 \bold{H}_{1,0},
\end{equation}
where
\begin{align}
\label{eq:def:D}
    \bold{D}_i & = \text{diag}(\boldsymbol{\alpha}^{(i)}), \quad   \boldsymbol{\alpha}^{(i)}  = [\alpha_{1}^{(i)}, \alpha_{2}^{(i)}, \ldots, \alpha_{m_i}^{(i)})]^T,
\end{align}
for each $i \in \{1,2, \ldots, n\}$. Here $\text{diag}(\boldsymbol{\alpha}^{(i)})$ is the diagonal matrix whose diagonal comprises $\boldsymbol{\alpha}^{(i)} \in (\mathbb{R}^+)^{m_{i}}$. 

We assume these channels are static and reciprocal during some period of consecutive transmission of symbols, where signals are received with noise and amplified in each layer. The system model for DL and UL transmission of a symbol $x$ is given by 
\begin{align}
\label{eq:main_sys_model}
 y  =~& h_{\text{tot}}x + w_{\text{DL}}, \\
 y  =~& h_{\text{tot}}^Tx + w_{\text{UL}},
 \end{align}
 respectively, and 
 \begin{align}
 w_{\text{DL}} =~& \sum\limits_{i=1}^{n}\left(\prod\limits_{j=i}^{n}\bold{H}_{j+1,j}\bold{D}_j\right)\bold{w}_{i}  +  w_{n+1},\\
 w_{\text{UL}} =~& \sum\limits_{i=1}^{n}\left(\prod\limits_{j=i}^{n}\bold{H}^{T}_{n+1 -j,n-j}\bold{D}_{n+1-j}\right)\bold{w}_{n+1-i}  +  w_{0}.
\end{align}
where $(\cdot)^T$ is the transpose. Here $y$ is the received symbol, and the receiver noise $\bold w_i~\sim~\mathcal{C}\mathcal{N}(\bold{0}_{m_i}, \sigma_{i}^2\bold{I}_{m_i})$ for each layer $i \in \{1,2, \ldots, n\}$ is assumed to be the same for DL and UL, where $\bold{0}_{m_i}$ is the vector of length $m_i$ with all elements equal to $0$ and $\sigma_i$ is the variance for the noise at layer $i$. $w_{n+1}~\sim \mathcal{C}\mathcal{N}(0, \sigma_{n+1}^2)$ and $ w_{0}~\sim \mathcal{C}\mathcal{N}(0, \sigma_{0}^2)$ is receiver noise at UE and BS, respectively.

\subsection{Signal-to-noise ratio (SNR)}
It holds that $w_{\text{DL}} \sim~\mathcal{C}\mathcal{N}(0, \sigma_{\text{DL}}^2)$ and $w_{\text{UL}} \sim~\mathcal{C}\mathcal{N}(0, \sigma_{\text{UL}}^2)$, where 
\begin{align}
    \label{eq:sigma_DL}
    \sigma_{\text{DL}}^2 & = \sum\limits_{i=1}^{n}E\Bigg(\Bigg |\left(\prod\limits_{j=i}^{n}\bold{H}_{j+1,j}\bold{D}_j\right)\bold{w}_{i}\bigg|^2\Bigg)  + \sigma_{n+1}^2, \\
    \label{eq:sigma_UL}
    \sigma_{\text{UL}}^2 & = \sum\limits_{i=1}^{n}E\Bigg(\bigg |\left(\prod\limits_{j=i}^{n}\bold{H}_{n+1 -j,n-j}\bold{D}_{n+1-j}\right)\bold{w}_{n+1-i}\bigg |^2\Bigg)  + \sigma_{0}^2.
\end{align}
The signal-to-noise ratio (SNR) is given by  
\begin{equation}
    \text{SNR}_{\text{DL}} = \frac{|h_{\text{tot}}|^2}{\sigma_{\text{DL}}^2} \quad \text{ and } \quad  \text{SNR}_{\text{UL}} = \frac{|h_{\text{tot}}|^2}{\sigma_{\text{UL}}^2},
\end{equation}
for DL-transmission and UL-transmission, respectively.


\section{Problem Formulation}
\label{subsec:problem_formulation}
The problem addressed is to choose, under constraints, the $\boldsymbol{\alpha}^{(i)}$'s in such a manner that $\text{SNR}_{\text{DL}}$ is maximized and $\text{SNR}_{\text{UL}}$ is maximized for DL-transmission and UL-transmission, respectively. However, the $\boldsymbol{\alpha}^{(i)}$'s, via the $\bold D_i$-matrices (see \eqref{eq:def:D}) appear nonlinearly in a product in the numerator and nonlinearly in a sum-product in denominator in both $\text{SNR}_{\text{DL}}$ and $\text{SNR}_{\text{UL}}$. Furthermore, since in general $\sigma_{\text{UL}} \neq \sigma_{\text{DL}}$, different solutions for DL- and UL-transmission is to be expected.  However, the numerator for both $\text{SNR}_{\text{DL}}$ and $\text{SNR}_{\text{UL}}$ is equal to $|h_{\text{tot}}|^2$.

The strategy we propose to get around this is to simply try to maximize the numerator $|h_{\text{tot}}|^2$ for $\text{SNR}_{\text{DL}}$ and $\text{SNR}_{\text{UL}}$. As we will show, such a strategy showcase significant improvement of the SNR, see Section~\ref{sec:num_res}, in comparison to expected SNR for random selection of the $\boldsymbol{\alpha}^{(i)}$'s, see Section~\ref{sec:snr_bounds}. 

Formally, the reduced problem addressed is on the form below.
\begin{align}      
\label{alg:1}
\mathop{\textnormal{maximize}}\limits_{\boldsymbol{\alpha}^{(i)} \in \mathcal{S}_i ~\forall~i}~|\boldsymbol{{h}}_{\text{tot}}|^2_2,
\end{align}
where the $\mathcal{S}_i$-sets are compact and ${h}_{\text{tot}}$ is a function of the ${\boldsymbol{\alpha}^{(i)}}'s$, see \eqref{eq:h_tot}. We refer to the $\mathcal{S}_i$-sets as activation sets and we propose different choices for such in Section~\ref{sec:activation}.


\section{The proposed method -- POLARNet}~\label{sec:algorithm}
To address problem \eqref{alg:1}, we propose Algorithm~\ref{alg:gen_alg} POLARNet, that monotonically improves the objective by selecting the $\boldsymbol{\alpha}^{(i)}$-amplifications layer-by-layer via local optimization. 

Now, for a diagonal matrix $\bold{D}$ with the vector $\bold{d}$ as its diagonal vector, we define $\text{diag}(\bold{D}) = \bold{d}$. Thus, the notation $\text{diag}(\cdot)$ may be used either for vector-inputs or matrix-inputs depending on the context. We further define for $n + 1 \geq i_2 > i_1 \geq 0$,
\begin{align}
    {\bold{{H}}}_{i_2, i_1}  & = \prod\limits_{i = i_1}^{i_2 - 1} \bold{H}_{i+1,i}\bold{D}_{i},
\end{align}
where $\bold{D}_0 = 1$. These are the channel matrices (or vectors) containing pairwise channels between layer $i_1$ and layer $i_2$ or layer $i_1$ and the UE or the BS and layer $i_2$. Matrix products are computed by left-multiplications. Now we observe that we can rewrite \eqref{eq:h_tot} as ${{h}}_{\text{tot}} = \bold{Y}_i\boldsymbol{\alpha}^{(i)}$
where
\begin{equation}\label{eq:D_i_0}
\bold{Y}_i =  {\bold{{H}}}_{n+1,i} {\bold{{D}}}_{i,0} \text{ and } {\bold{D}}_{i,0} = \text{diag}({\bold{H}}_{i,0}).
\end{equation}

\subsection{Algorithm}
\begin{algorithm}[H]
    \caption{POLARNet}\label{alg:gen_alg} 
    \vspace*{0.08cm} 
    \noindent
    \textbf{Input:} 
    channel matrices $\{\bold{H}_{i+1,i}\}_{i = 0}^n$ initial repeater amplifications $\{\boldsymbol{\alpha}^{(i)}\}_{i =1}^n$  \\
    \noindent
    \textbf{Output:} locally  optimized repeater amplifications $\{\boldsymbol{\alpha}^{(i)}\}_{i = 1}^n$
    \vspace*{1mm}
    \hrule 
    \begin{algorithmic}[1]
    \vspace*{0.08cm}
        \STATE \textbf{repeat}
        \STATE \hspace{0.5cm} \textbf{for each} $i \in \{1, 2, \ldots, n\}$ \textbf{do }
        \STATE \hspace{1.0cm} compute $\bold Y_i$
        \STATE \hspace{1.0cm} $\boldsymbol{\alpha}^{(i)} \gets \argmax\limits_{\bold{\alpha} \in \mathcal{S}_i} \boldsymbol{\alpha}^{(i)T} \text{Re}\{\bold{Y}_i^H\bold{Y}_i\}\boldsymbol{\alpha} $
        \STATE \hspace{0.5cm} \textbf{end for}
        \STATE \textbf{until} convergence
    \end{algorithmic}
\end{algorithm}

In Algorithm~\ref{alg:gen_alg}, $\boldsymbol{\alpha}^{(i)T}$ is the transpose of $\boldsymbol{\alpha}^{(i)}$. The convergence criterion at line 6, could for example be given as finite number of repetitions $N$ for the outer loop, so that the total amount of iterations (for the inner loop) becomes $Nn$. The convergence criterion could also be formulated in terms of the improvement of the objective function, \emph{e.g.}, the algorithm terminates if $|{h}_{\text{tot}}|_2$ has not increased more than $\epsilon \geq 0$ during $n$ iterations. 

\subsection{Monotone convergence of the objective}~\label{sec:monotone}
The objective in \eqref{alg:1} increases monotonically under Algorithm~\ref{alg:gen_alg}. To show this, let $\boldsymbol{\alpha}_{\text{old}} = \boldsymbol{\alpha}^{(i)}$ and $\boldsymbol{\alpha}_{\text{new}} \in \argmax\limits_{\bold{\alpha} \in \mathcal{S}_i} \boldsymbol{\alpha}^{(i)T} \bold{Y}_i^H\bold{Y}_i\boldsymbol{\alpha}$, where $i \in \{1,2, \ldots, n\}$. 

For convenience, let $\bold W = \bold{Y}_i^H\bold{Y}_i$. We  define $h^2_{\text{tot, old}} = \boldsymbol{\alpha}_{\text{old}} \bold W \boldsymbol{\alpha}_{\text{old}}$ and $h^2_{\text{tot},\text{new}} = \boldsymbol{\alpha}_{\text{new}} \bold W \boldsymbol{\alpha}_{\text{new}}$, which are is the objective value before and after $\boldsymbol{\alpha}^{(i)}$ is updated, respectively. 

Since $\boldsymbol{\alpha}_{\text{old}}$ and $\boldsymbol{\alpha}_{\text{new}}$ are real, 
\begin{align}
    \nonumber
    & 0 \leq (\boldsymbol{\alpha}_{\text{old}} - \boldsymbol{\alpha}_{\text{new}})^T \bold W (\boldsymbol{\alpha}_{\text{old}} - \boldsymbol{\alpha}_{\text{new}}) \\
    \nonumber
    =~ & \boldsymbol{\alpha}_{\text{old}}^T \text{Re}\{\bold W\} \boldsymbol{\alpha}_{\text{old}} - 2 \boldsymbol{\alpha}_{\text{old}}^T \text{Re}\{\bold W\} \boldsymbol{\alpha}_{\text{new}} + \boldsymbol{\alpha}_{\text{new}} \text{Re}\{\bold W\} \boldsymbol{\alpha}_{\text{new}} \\
    \nonumber
     \leq~& \boldsymbol{\alpha}_{\text{new}} \text{Re}\{\bold W\} \boldsymbol{\alpha}_{\text{new}} - \boldsymbol{\alpha}_{\text{old}}^T \text{Re}\{\bold W\} \boldsymbol{\alpha}_{\text{old}} \\
     =~& \boldsymbol{\alpha}_{\text{new}} \bold W \boldsymbol{\alpha}_{\text{new}} - \boldsymbol{\alpha}_{\text{old}}^T \bold W \boldsymbol{\alpha}_{\text{old}} = h^2_{\text{tot, new}} - h^2_{\text{tot, old}}. 
\end{align}

\subsection{Activation sets}~\label{sec:activation}
In this section we introduce some different choices of activation sets (i.e., the compact $\mathcal{S}_i$-sets) and explain what they represent. We provide, for each activation set introduced, an optimization procedure for solving the maximization problem at line 4 in Algorithm~\ref{alg:gen_alg}. In what follows, let $\beta_i > 0$ for $i \in \{1,2, \ldots, n\}$.

\subsubsection{The non-negative $2$-ball}~\\
{Activation set:} $\mathcal{S}_i = \{\boldsymbol \alpha \in (\mathbb{R}^+)^{m_i}: \|\boldsymbol{\alpha}\|_2 \leq \beta_i\}$.
{Explanation:}
Total power used for the repeaters in layer $i$ is less or equal to $\beta_i^2$. {Optimization procedure:}
\begin{equation}
    \boldsymbol{\alpha}^{(i)} \gets \beta_i\frac{\text{ReLU}(\text{Re}\{\bold{Y}_i^H\bold{Y}_i\}\boldsymbol{\alpha}^{(i)})}{\|\text{ReLU}(\text{Re}\{\bold{Y}_i^H\bold{Y}_i\}\boldsymbol{\alpha}^{(i)})\|_2},
\end{equation}
where $\text{ReLU}(\bold x) = \max\{\bold{0}, \bold{x}\}$ is done element-wise for a real vector $\bold x$. Positive elements in the vector $\text{Re}\{\bold{Y}_i^H\bold{Y}_i\}\boldsymbol{\alpha}$ exist provided $\|\bold{h}_{\text{tot}}\|_2^2$ initially is strictly larger than $0$ (which is assumed throughout). This is an important activation set due to the following reason. Suppose we want to solve the following problem 
\begin{align}      
\label{alg:alternative:1}
\mathop{\textnormal{maximize} }~|\boldsymbol{\text{h}}_{\text{tot}}|^2_2, \text{ s.t. } \sum_{i = 1}^n(\boldsymbol{\alpha}^{(i)})^2 = \beta^2 > 0.
\end{align}
Then an optimal solution to \eqref{alg:alternative:1} is obtained by solving \eqref{alg:1} with $\mathcal{S}_i = \{\boldsymbol \alpha \in (\mathbb{R}^+)^{m_i}: \|\boldsymbol{\alpha}\|_2 \leq \beta_i\}$ for each $i$, where $\sum_{i = 1}^n\beta_i^2 = \beta^2$.  

\subsubsection{The non-negative $\infty$-ball}~\\
{Activation set:} $\mathcal{S}_i = \{\boldsymbol \alpha \in (\mathbb{R}^+)^{m_i}: \|\boldsymbol{\alpha}\|_{\infty} \leq \beta_i\}$. {Explanation:} The power used for each repeater in layer $i$ is not larger than $\beta_i^2$. {Optimization procedure:}
\begin{equation}
    \boldsymbol{\alpha}^{(i)} \gets \beta_i\text{sgn}^+(\text{Re}\{\bold{Y}_i^H\bold{Y}_i\}\boldsymbol{\alpha}^{(i)}),
\end{equation}
where $\text{sgn}^+(x)$ is equal to $1$ if $x > 0$ and otherwise $0$. $\text{sgn}^+(\bold x)$ is the element-wise extension for a real vector $\bold x$. 

\subsubsection{Select at most K policy}~\\
{Activation set:} $\mathcal{S}_i = \{\boldsymbol \alpha = [\alpha_1, \alpha_2, \ldots, \alpha_{m_i}] \in \{0, \beta_i\}^{m_i}: \sum_{j = 1}^{m_i}\alpha_i \leq  K\beta_i\}$. 
{Explanation:} Select at most $K$ repeaters as active in layer $i$, each with power $\beta_i^2$. All other repeaters are inactive.
{Optimization procedure:}
\begin{equation}
    \boldsymbol{\alpha}^{(i)} \gets \beta_i \text{maxK}^+(\text{Re}\{\bold{Y}_i^H\bold{Y}_i\}\boldsymbol{\alpha}^{(i)}),
\end{equation}
where, for a real vector $\bold x$ with at least one positive element, $\text{maxK}^+(\bold x)$ is a binary vector. If $\bold x$ contains $N_{\text{pos}}$ positive elements, $\text{maxK}^+(\bold x)$ contains $\min\{N_{\text{pos}}, K\}$ elements equal to $1$, which correspond to the $\min\{N_{\text{pos}}, K\}$ largest positive elements in $\bold x$; all other elements in $\text{maxK}^+(\bold x)$ are $0$. 

\subsubsection{The non-negative $1$-ball or select one policy}~\\\label{subsubsec:select1}
{Activation set:} $\mathcal{S}_i = \{\boldsymbol \alpha \in (\mathbb{R}^+)^{m_i}: \|\boldsymbol{\alpha}\|_1 \leq \beta_i\}$.
{Explanation:} An optimal solution is to select, for layer $i$, only one repeater as active with power $\beta_i^2$. This activation set is a special case of 3) above, where $K = 1$.
{Optimization procedure:} 
\begin{equation}
    \boldsymbol{\alpha}^{(i)} \gets \beta_i \text{max1}^+(\text{Re}\{\bold{Y}_i^H\bold{Y}_i\}\boldsymbol{\alpha}^{(i)}),
\end{equation}
where, for a real vector $\bold x$ with at least one positive element, $\text{max1}^+(\bold x)$ is a binary vector. The element in $\text{max1}^+(\bold x)$ corresponding to the largest positive element in $\bold x$ is equal to $1$ and all other elements in $\text{max1}^+(\bold x)$ are equal to $0$. It is a somewhat trivial activation set, where only one repeater is active in each layer, but included here as a separate activate set due to the following reason. If $j_i$ is the index of the element in $\boldsymbol{\alpha}^{(i)}$ that is equal to $\beta_i = 1$ (corresponding to the active repeater), we may rewrite $h_{\text{tot}}$ as 
\begin{align}
   |h_{\text{tot}}|  =~& |\bold{H}_{n+1,n}(j_i)|\left( \prod_{i= i}^{n-1}|\bold{H}_{i+1,i}(j_{i+1},j_i)|\right)|\bold{H}_{1,0}(j_1)|.
\end{align}
Maximization of $|h_{\text{tot}}|$ is equivalent to maximization of $\log(|h_{\text{tot}}|)$ over the sequence $\{j_i\}_{i = 1}^n$, which, when treated as longest path problem for our DAG-graph with $n$ layers, is solved in $\mathcal{O}(\sum_{i = 1}^{n-1}m_{i}m_{i+1})$ computational time. In Section~\ref{sec:num_res} we compare this optimal solution to solutions found when our algorithm is used for different activation sets. \\

\noindent 
If, for each layer $i$, the updating of the $\boldsymbol{\alpha}^{(i)}$'s is done according to one of the optimization procedures for the activation sets 1), 2), and 4), the objective value will converge after a finite number of iterations. For the activation set 1)-3), optimal points are on a sphere-boundary. Hence, we may refer to spheres instead of balls.

\subsection{Forward-backward-traversing layers}\label{sec:for_back}
In Algorithm~\ref{alg:gen_alg}, for a well-defined convergence criterion (at line 6), all layers will be traversed in some order (see line 2) $N < \infty$ times. Here we propose to traverse the layers forwards ($i = 1$, $i = 2$, \ldots, $i = n$) and backwards ($i = n$, $i = n-1$, \ldots, $i = 1$) alternately. Such a procedure allows for efficient recursive computation of the $\bold Y_i$'s as described below. The forward pass/traverse of layers is in the DL-direction, whereas the backward pass is in the UL-direction. 

{Forward-traversing layers:} Assumption: $\bold{H}_{n+1,i}$ is available for each layer $i$ and $\bold H_{1,0}$ is available for layer 1. For $i \geq 2$, compute $\bold H_{i,0} = \bold{H}_{i,i-1}\bold D_{i-1}\bold h_{i-1,0}$, and then compute $\bold Y_i$ according to \eqref{eq:D_i_0}. 

{Backward-traversing layers:} Assumption:  $\bold{D}_{i,0}$ (or equivalently $\bold H_{i,0}$) is available for each layer $i$ and $\bold H_{n+1,n}$ is available for layer $n$. For $i \leq n-1$, compute $\bold H_{n+1,i} =  \bold{H}_{n+1,i+1}\bold D_{i+1}\bold{H}_{i+1,i}$, and then compute $\bold Y_i$ according to \eqref{eq:D_i_0}. 

\subsubsection{Computational complexity}
The total computational complexity of the algorithm depends on the convergence criterion, which specifies $N$. Here we limit the analysis to the per-layer computational complexity; lines 3 and 4 of Algorithm~\ref{alg:gen_alg}. 
If the forward-backward procedure is used, see Section~\ref{sec:for_back}, the computational complexity for computing $\bold Y_i$ is $\mathcal{O}(m_{i-1}m_i)$ in the forward pass and $\mathcal{O}(m_{i+1}m_{i})$ in the backward pass. 

We see that the optimization procedures for the activation sets 1)-4) all operate on the vector $\text{Re}\{\bold{Y}_i^H\bold{Y}_i\}\boldsymbol{\alpha}^{(i)} = \text{Re}\{\bold{Y}_i^H\bold{Y}_i\boldsymbol{\alpha}^{(i)}\}$. The computational complexity for computing this vector given $\bold Y_i$,  is $\mathcal{O}(m_i)$. 

Given $\text{Re}\{\bold{Y}_i^H \bold{Y}_i \boldsymbol{\alpha}^{(i)}\} = \text{Re}\{\bold{Y}_i^H \bold{Y}_i \}\boldsymbol{\alpha}^{(i)}$, the optimization procedures for activation sets 1)-4), have computational complexities as follows. For activation sets 1), 2) and 4) the computational complexity is $\mathcal{O}(m_i)$ and for activation set 3) the computational complexity is $\mathcal{O}(m_i\log(K))$.


\subsection{SNR bounds}\label{sec:snr_bounds}
In this section we assume that for each layer $i$, the activation set $\mathcal{S}_i$ is chosen as either 1), 2), 3) or 4).
A common conservative upper bound on $\text{SNR}_{\text{DL}}$ and $\text{SNR}_{\text{UL}}$ may be constructed as follows. 
Let $\sigma = \min\{\sigma_{0}, \sigma_{n+1}\}$, whereby it follows, see  that $\sigma_{\text{DL}}$ and $\sigma_{\text{UL}}$ are both smaller than or equal to $\sigma$. Hence, 
\begin{equation}
    \text{SNR}_{\text{DL}},  \text{SNR}_{\text{UL}} \leq \frac{|h_{\text{tot}}|^2}{\sigma^2},
\end{equation}
i.e., both $\text{SNR}_{\text{DL}}$ and $\text{SNR}_{\text{UL}}$ is smaller or equal to $\frac{|h_{\text{tot}}|^2}{\sigma^2}$. 

Now, if we treat $\boldsymbol{\alpha}_{\text{all}} = [\boldsymbol{\alpha}^{(1)T}, \boldsymbol{\alpha}^{(2)T}, \ldots, \boldsymbol{\alpha}^{(n)T}]^T$ as a random vector, we define by $\text{E}_{\boldsymbol{\alpha}_{\text{all}}}(\text{SNR}_{\text{DL}})$ and $\text{E}_{\boldsymbol{\alpha}_{\text{all}}}(\text{SNR}_{\text{DL}})$ the expected SNR w.r.t. $\boldsymbol{\alpha}_{\text{tot}}$ for DL and UL, respectively. If we treat $\bold{h}_{\text{all}} = [\text{vec}(\bold{H}_{1,0})^T, \text{vec}(\bold{H}_{2,1})^T, \ldots, \text{vec}(\bold{H}_{n+1,n})^T]^T$, where $\text{vec}(\cdot)$ is the vectorization operator, as a random vector, we may analogously define $\text{E}_{\boldsymbol{h}_{\text{all}}}(\text{SNR}_{\text{DL}})$ and $\text{E}_{\boldsymbol{h}_{\text{all}}}(\text{SNR}_{\text{DL}})$. Finally the joint expected SNR is $\text{E}_{\boldsymbol{\alpha}_{\text{all}}, \boldsymbol{h}_{\text{all}}}(\text{SNR}_{\text{DL}})$ and $\text{E}_{\boldsymbol{\alpha}_{\text{all}},\boldsymbol{h}_{\text{all}}}(\text{SNR}_{\text{DL}})$ for DL and UL, respectively. It holds that
\begin{equation}\label{eq:SNR_upper}
    \text{E}_{\boldsymbol{\alpha}_{\text{all}}, \boldsymbol{h}_{\text{all}}}(\text{SNR}_{\text{DL}}),   \text{E}_{\boldsymbol{\alpha}_{\text{all}}, \boldsymbol{h}_{\text{all}}}(\text{SNR}_{\text{UL}}) \leq  \frac{\text{E}_{\boldsymbol{\alpha}_{\text{all}}, \boldsymbol{h}_{\text{all}}}(|h_{\text{tot}}|^2)}{\sigma^2},
\end{equation}
since $\sigma_{\text{UL}}$ and $\sigma_{\text{DL}}$ are lower-bounded by $\sigma^2$ for all realizations of $\boldsymbol{\alpha}_{\text{all}}$ and $\boldsymbol{h}_{\text{all}}$. 

We now assume that the elements in $\boldsymbol{h}_{\text{all}}$ are IID with distribution $\mathcal{C}\mathcal{N}(0, \sigma_{H}^2)$, and assume one of the three types of distributions below for the $\boldsymbol{\alpha}^{(i)}$-vectors (which are relevant for the activation sets previously described): 1) each $\alpha^{(i)}$-vector is uniformly distributed on non-negative orthant of the euclidean sphere; 2) each $\alpha^{(i)}$-vector is a uniformly distributed one-hot vector where one element is equal to 1 and the rest are equal to 0;  each $\alpha^{(i)}$-vector have elements that are either $1$ or $0$, with $0.5$ probability for either choice.

Under these circumstances,  $\text{E}_{\boldsymbol{\alpha}_{\text{all}}, \boldsymbol{h}_{\text{all}}}(|h_{\text{tot}}|^2)$ can be written as $E_{\bold{H}_{1,0}}(\bold{H}_{1,0}^H \cdots E_{\bold{D}_n}(\bold{D}_nE_{\bold{H}_{n+1,n}}(\bold{H}^H_{n+1,n} \bold{H}_{n+1,n})\bold{D}_n) \\ \cdots)\bold{H}_{1,0})$ and this expression evaluates to $\sigma_H^{n+1}$ for distributions 1) and 2) for the $\boldsymbol{\alpha}^{(i)}$-vectors and it evaluates to $\frac{(m_1\cdots m_n)\sigma_H^{n+1}}{4^n}$ for distribution 3). By using \eqref{eq:SNR_upper}, we may conclude that for distributions 1) and 2) \eqref{eq:SNR_upper:2} holds below, and for distribution 3) \eqref{eq:SNR_upper:3} holds below.
\begin{align}\label{eq:SNR_upper:2}
    & \text{E}_{\boldsymbol{\alpha}_{\text{all}}, \boldsymbol{h}_{\text{all}}}(\text{SNR}_{\text{DL}}),   \text{E}_{\boldsymbol{\alpha}_{\text{all}}, \boldsymbol{h}_{\text{all}}}(\text{SNR}_{\text{UL}}) \leq  \frac{\sigma_H^{n+1}}{\sigma^2}, \\
    \label{eq:SNR_upper:3}
    & \text{E}_{\boldsymbol{\alpha}_{\text{all}}, \boldsymbol{h}_{\text{all}}}(\text{SNR}_{\text{DL}}),   \text{E}_{\boldsymbol{\alpha}_{\text{all}}, \boldsymbol{h}_{\text{all}}}(\text{SNR}_{\text{UL}}) \leq  \frac{(m_1\cdots m_n)\sigma_H^{n+1}}{4^n \sigma^2}.
\end{align}


\begin{figure*}[!t]
    \centering
    \begin{minipage}{0.45\linewidth}
        \centering
        \includegraphics[width=\linewidth]{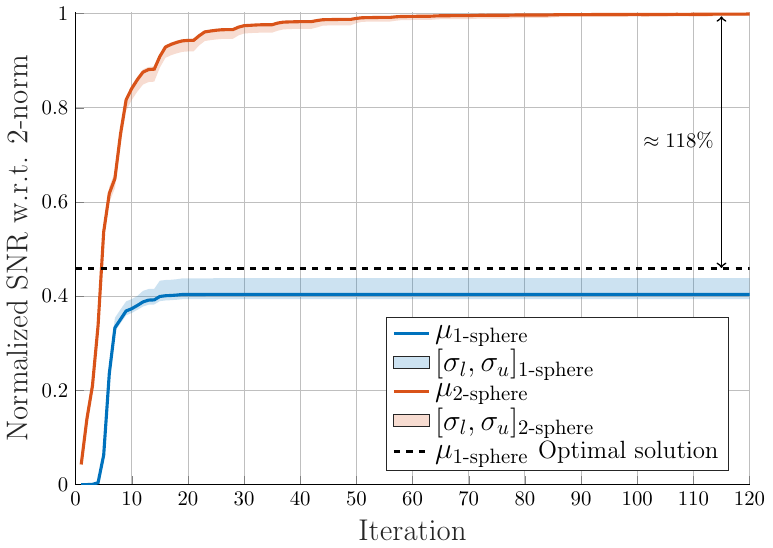}
        \caption{Convergence performance for the considered scenario and comparison to the optimal solution in Section~\ref{subsubsec:select1}.}
        \label{fig:polarnet_optsol}
    \end{minipage}%
    \hspace{30pt}
    \begin{minipage}{0.45\linewidth}
        \centering
        \includegraphics[width=\linewidth]{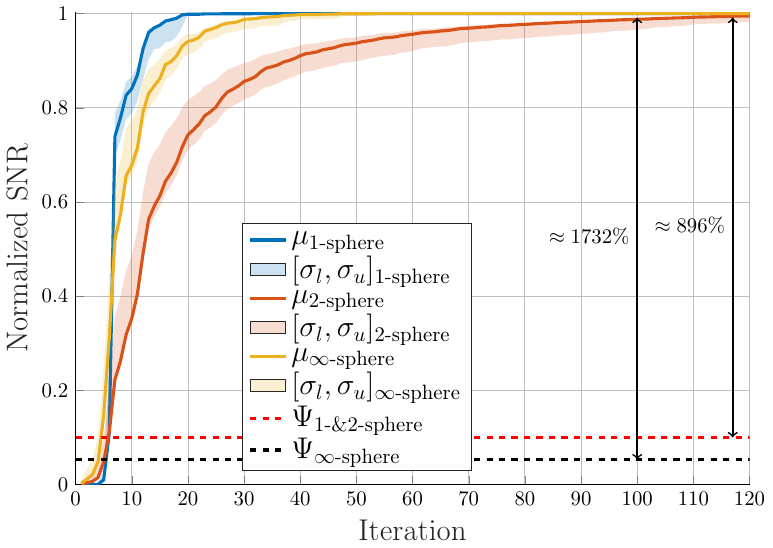}
        \caption{Convergence performance for the scenario in Section~\ref{sec:snr_bounds} and comparison to the bounds on SNR in \eqref{eq:SNR_upper:2}, \eqref{eq:SNR_upper:3}.}
        \label{fig:polarnet_snrbounds}
    \end{minipage}
    \label{fig:polarnet_combined}
\end{figure*}

\section{Numerical Results}
\label{sec:num_res}
In this section, we numerically evaluate Algorithm~\ref{alg:gen_alg}  under Rician fading ($K=0.5$) for the three types of activation sets, i.e., $\mathcal{S}_i$-sets, 1), 2), and 4), see Section~\ref{sec:activation}.  We verify monotone convergence, compare performance to the optimal solution for activation set 4) in Section~\ref{subsubsec:select1}  and also compare performance to the $\text{SNR}$ bounds in Section~\ref{sec:snr_bounds}.

In the first scenario repeaters are deployed on a grid with $n=7$ layers, interlayer separation $100$\,m and intralayer separation $10$\,m (between repeaters within each layer) with $m_i=\{6,13,4,5,11,8,7\}$ repeaters in layer $i=1,2, \ldots, 7$. Dominant-path amplitudes undergo free-space path loss $\gamma(d) = \lambda^2/(4 \pi d)^2$ ($\lambda$ corresponds to $2$\, GHz carrier). Scattered paths are modeled as complex Gaussian noise. The $\boldsymbol{\alpha}^{(i)}$'s are initialized by creating a vector where elements are drawn from the uniform distribution with $(0,1)$, which is subsequently normalized w.r.t.\ the $2$-norm and the $1$-norm for activation sets 1) and 4), respectively. For each activation set, $10^4$ independent experiments are conducted. In each experiment, Algorithm~\ref{alg:gen_alg} performs $N=20$ outer-loop iterations (in total, there are $Nn$ iterations). The objective value at each iteration is normalized to $[0, 1]$, where $1$ corresponds to the final objective value for activation set 1). For the $10^4$ experiments, we compute the mean $\mu$ of the normalized objectives, together with the asymmetric standard deviations: $\mathcal{T}_u = \{x \mid x \geq \mu \}$ and $\mathcal{T}_l = \{x \mid x < \mu \}$; then $\sigma_u=\sum_{x\in \mathcal{T}_u} \frac{\mu - x}{Nn}$ and $\sigma_l=\sum_{x\in \mathcal{T}_l} \frac{\mu - x}{Nn}$.

Results for this scenario is shown in Fig.~\ref{fig:polarnet_optsol}. Besides fast convergence, the algorithm with activation set 4) is quite close to the optimal solution for activation set 1) and the algorithm with activation set 1) is superior (approximately $118$\% better) to the optimal solution for activation set 4) for the same power utilization.

 In Fig.~\ref{fig:polarnet_snrbounds} we present results for a modification of the first scenario where no dominant path components are present and the elements of the channel matrices are IID and $\mathcal{C}{N}(0, \sigma_{H}^2)$. Significant improvement w.r.t. upper bounds on expected SNR $\Psi_{1\text{-} \& 2\text{-sphere}}$ (896\%) and $\Psi_{\infty\text{-sphere}}$ (1732\%), corresponding to the right-hand side of \eqref{eq:SNR_upper:2} and \eqref{eq:SNR_upper:3}, respectively, can be observed.

\section{Conclusion}
A novel method, POLARNet, was presented for computationally fast selection of power in multi-layer repeater networks. The method is gradient-free, step-size-free, and has proven objective-monotonicity. Numerical simulations show significant improvement w.r.t. upper bounds on expected SNR. Additionally, power distribution over multiple repeaters is shown to be superior to optimal selection of single repeaters in the layers.

\bibliographystyle{ieeetr}
\bibliography{refs.bib}

@ARTICLE{spectral_eff_D_mimo,
  author={Wang, Dongming and Wang, Jiangzhou and You, Xiaohu and Wang, Yan and Chen, Ming and Hou, Xiaoyun},
  journal={IEEE Journal on Selected Areas in Communications}, 
  title={Spectral Efficiency of Distributed MIMO Systems}, 
  year={2013},
  volume={31},
  number={10},
  pages={2112-2127},
  doi={10.1109/JSAC.2013.131012}}

@inproceedings{larsson_stability,
  title={Stability analysis of interacting wireless repeaters},
  author={Larsson, Erik G and Bai, Jianan},
  booktitle={2024 IEEE 25th International Workshop on Signal Processing Advances in Wireless Communications (SPAWC)},
  pages={756--760},
  year={2024},
  organization={IEEE}
}

@ARTICLE{saras_paper,
  author={Willhammar, Sara and Iimori, Hiroki and Vieira, Joao and Sundström, Lars and Tufvesson, Fredrik and Larsson, Erik G.},
  journal={IEEE Communications Magazine}, 
  title={Achieving Distributed MIMO Performance with Repeater-Assisted Cellular Massive MIMO}, 
  year={2025},
  volume={63},
  number={3},
  pages={114-119},
  doi={10.1109/MCOM.001.2400332}}

@manual{3gpp_ncr,
    organization = {},
    title = {3GPP TS 38.106 version 18.3.0, NR repeater radio transmission and reception (Realease 18)},
    number = {Std},
    year = {2024},
    month = {Jan.},
    note = {}
}

@article{topal2025fair,
  title={Fair and Energy-Efficient Activation Control Mechanisms for Repeater-Assisted Massive MIMO},
  author={Topal, Ozan Alp and Demir, {\"O}zlem Tu{\u{g}}fe and Bj{\"o}rnson, Emil and Cavdar, Cicek},
  journal={arXiv preprint arXiv:2504.03428},
  year={2025}
}

@article{wen2024shaping,
  title={Shaping a smarter electromagnetic landscape: IAB, NCR, and RIS in 5G standard and future 6G},
  author={Wen, Chao-Kai and Tsai, Lung-Sheng and Shojaeifard, Arman and Liao, Pei-Kai and Wong, Kai-Kit and Chae, Chan-Byoung},
  journal={IEEE Communications Standards Magazine},
  volume={8},
  number={1},
  pages={72--78},
  year={2024},
  publisher={IEEE}
}

@article{tsai2010capacity,
  title={Capacity scaling and coverage for repeater-aided MIMO systems in line-of-sight environments},
  author={Tsai, Lung-Sheng and Shiu, Da-shan},
  journal={IEEE transactions on wireless communications},
  volume={9},
  number={5},
  pages={1617--1627},
  year={2010},
  publisher={IEEE}
}

@article{sanguinetti2012tutorial,
  title={A tutorial on the optimization of amplify-and-forward MIMO relay systems},
  author={Sanguinetti, Luca and D'Amico, Antonio A and Rong, Yue},
  journal={IEEE Journal on Selected Areas in Communications},
  volume={30},
  number={8},
  pages={1331--1346},
  year={2012},
  publisher={IEEE}
}

@inproceedings{haliloglu2023distributed,
  title={Distributed MIMO systems for 6G},
  author={Haliloglu, Omer and Yu, Han and Madapatha, Charitha and Guo, Hao and Kadan, Fehmi Emre and Wolfgang, Andreas and Puerta, Rafael and Frenger, P{\aa}l and Svensson, Tommy},
  booktitle={2023 Joint European Conference on Networks and Communications \& 6G Summit (EuCNC/6G Summit)},
  pages={156--161},
  year={2023},
  organization={IEEE}
}

@article{hu2018beyond,
  title={Beyond massive MIMO: The potential of data transmission with large intelligent surfaces},
  author={Hu, Sha and Rusek, Fredrik and Edfors, Ove},
  journal={IEEE Transactions on Signal Processing},
  volume={66},
  number={10},
  pages={2746--2758},
  year={2018},
  publisher={IEEE}
}

@article{rumelhart1986learning,
  title={Learning representations by back-propagating errors},
  author={Rumelhart, David E and Hinton, Geoffrey E and Williams, Ronald J},
  journal={nature},
  volume={323},
  number={6088},
  pages={533--536},
  year={1986},
  publisher={Nature Publishing Group UK London}
}

@article{liu2024survey,
  title={A survey of recent advances in optimization methods for wireless communications},
  author={Liu, Ya-Feng and Chang, Tsung-Hui and Hong, Mingyi and Wu, Zheyu and So, Anthony Man-Cho and Jorswieck, Eduard A and Yu, Wei},
  journal={IEEE Journal on Selected Areas in Communications},
  year={2024},
  publisher={IEEE}
}

@inproceedings{schumacher2022increasing,
  title={Increasing cellular network energy efficiency for railway corridors},
  author={Schumacher, Adrian and Merz, Ruben and Burg, Andreas},
  booktitle={2022 Design, Automation \& Test in Europe Conference \& Exhibition (DATE)},
  pages={1103--1106},
  year={2022},
  organization={IEEE}
}

\end{document}